\documentclass[aps,prx,twocolumn,notitlepage,amsmath,amstex,amssymb,superscriptaddress]{revtex4-1}

\usepackage{bm}
\usepackage{lmodern} 
\usepackage{microtype} 
\usepackage[english]{babel}

\usepackage{color,graphicx,comment}

\usepackage{dsfont,url}
\usepackage{slashed,cancel}
\usepackage[usenames,dvipsnames]{xcolor}
\usepackage[colorlinks=true, linkcolor=black, citecolor=ForestGreen]{hyperref}
\usepackage{mdframed}

\usepackage{physics,tikz}

\usepackage[caption=false]{subfig}

\usetikzlibrary{decorations.markings,calc}

\begin{document}

\title{\bf 
Quantum Critical Solids
}

\author{Subham Dutta Chowdhury}
\affiliation{The Abdus Salam ICTP, Strada Costiera 11, 34151, Trieste, Italy}
\author{Leo Radzihovsky}
\affiliation{Department of Physics and Center for Theory of Quantum Matter,
University of Colorado, Boulder CO 80309, USA}
\author{Luca V. Delacr\'etaz}
\affiliation{Leinweber Institute for Theoretical Physics \& James Franck Institute, University of Chicago, Chicago, IL 60637, USA}

\begin{abstract}
We study the fate of quantum solids with a vanishing velocity of transverse phonons. Such ``floppy'' solids exhibit a cubic elastic nonlinearity that is relevant in spatial dimensions $d<4$ and drives the system to a strongly-coupled Lifshitz quantum critical point in $d=3$, that we analyze in an $\epsilon = 4-d$ expansion. The associated phase transition is generically first-order, preempting the system's full access to the strongly-coupled fixed point. Nevertheless, the critical data of this quantum critical point is accessible within a Ginzburg region in the vicinity of the transition. This cubic nonlinearity  vanishes exactly in $d=2$, where the leading interactions are instead quartic and are marginal, leading to weak logarithmic singularities. We discuss possible realizations of this ``crystal Lifshitz criticality'' in quantum materials.

\end{abstract}

\maketitle

\section{Introduction and Motivation}

Many gapless phases of matter, including superfluids, solids, antiferromagnets, and Fermi liquids, can be captured by simple and universal effective field theories (EFTs). 
Because EFTs are largely insensitive to underlying microscopic details, they serve as a powerful common language across subfields of physics, with similar EFTs describing the relativistic vacuum of particle physics, early universe cosmology, and emergence in soft matter and quantum many-body physics.
However, the EFTs that describe the gapless phases of condensed matter share a peculiar feature that does not exist around the relativistic vacuum:
the existence of exactly marginal tunable parameters, such as the velocities of phonons, or the Landau parameters of Fermi liquids. This naturally raises the question of whether interesting new phenomena can arise in limits of this parameter space.

In the simplest EFT of a single (scalar) Nambu-Goldstone field (e.g., a superfluid), tuning its velocity to zero leads to a  Lifshitz quantum critical point (QCP) that becomes strongly coupled in spatial dimensions $d<2$ \cite{Yang:2004ferr,Kozii:2017ferr}. This {\em scalar} Lifshitz QCP has found numerous applications across fields: in $d=3$, where it remains weakly coupled, it has served as a model of cosmological inflation \cite{Arkani-Hamed:2003juy,Cheung:2007st}; in $d=2$, it arises as a deconfined quantum critical point in quantum spin liquids \cite{Vishwanath:2003yjl,Ardonne:2003wa,Fradkin:2003vaq}; finally in $d=1$, where it is strongly coupled and perhaps most interesting, it describes the fate of an Ising model coupled to a Luttinger liquid (which can model, for example, an itinerant ferromagnetic quantum phase transition) 
\cite{Yang:2004ferr,Kozii:2017ferr,Cole_2019}. Numerous closely related observations in the literature include \cite{PhysRevLett.47.856,PhysRevA.26.2196,Sachdev:1996zero,Zedler:2005ferr,ffloLR,Radzihovsky:2011zz,criticalMatterLR,Nahum:2025bdg,Antunes:2026jjf}.

In this Letter, we study nontrivial behavior of quantum solids, with a shear modulus or equivalently transverse phonon velocity tuned to zero. As we demonstrate, such a ``floppy'' or anomalously soft solid is in fact quantum critical, with its phenomenology controlled by a strongly-coupled QCP, which we refer to as ``{\em crystal} Lifshitz'' universality. While one may expect a similar fate to the aforementioned scalar Lifshitz universality, the low-energy dynamics of floppy solids is in fact qualitatively different: the spatial-vector nature of phonons allows for {\em cubic} (rather than quartic) nonlinearities, leading to larger quantum fluctuations. As a result, floppy quantum solids are described by a strongly coupled QCP even in $d=3$, which we study in a controlled $\epsilon=4-d$ expansion.

The same cubic nonlinearities that enhance quantum fluctuations also have the less appealing feature that they typically drive the transition to first-order. However, we argue below that for a weakly first-order transition the data of the QCP can still be visible in $d=3$ quantum solids with slow phonons: a Ginzburg criterion shows that a small nonzero phonon velocity can stabilize the critical solid without washing away its universal critical scaling. 

In contrast, we find that the dynamics of floppy quantum solids is entirely different in $d=2$, because the cubic nonlinearity between transverse phonons vanishes identically. In this case, the leading nonlinearities are quartic, similar to the scalar Lifshitz dynamics, but with a richer landscape of possible IR fates.

Lifshitz criticality has a long history in soft matter physics (see Ref.~\cite{criticalMatterLR} for a review). Realizations in quantum many-body systems have attracted interest more recently, from quantum dimer models and deconfined criticality \cite{Vishwanath:2003yjl,Ardonne:2003wa,Fradkin:2003vaq} to systems with dipole conservation \cite{Lake:2022ico,Zechmann:2022esq,Anakru:2023jcg,Han:2024mzk}. The quantum setting also brings qualitatively new ingredients: Lifshitz dynamics of fermions \cite{Son:2007ja}, topological Berry phase terms \cite{Anakru:2024jec}, and a different counting of dimensions that leaves room for new fixed points, such as the strongly coupled one we discuss below.

\section{Elasticity theory with slow phonons: scaling argument}

The effective field theory of elasticity describes a vector degree of freedom $u_i(t,x)$, the Nambu-Goldstone boson (or phonon) associated with the spontaneous breaking of translations. In the simplest case of isotropic solids, its action $S =\int dt d^dx\ {\cal L}$ takes the general form
\begin{align}\label{eq_L}
{\cal L} &= \frac{\chi}2(\partial_t u_i)^2 
 - V_{\rm el}(u_{ij})
  - \frac12 K (\nabla^2 u_i)^2
  + \cdots\, , 
\end{align}
where $u_{ij} = \tfrac12(\partial_iu_j+\partial_ju_i +\partial_iu_k\partial_ju_k)$ is the nonlinear strain tensor, $\chi$ is the momentum susceptibility, and $V_{\rm el}$ the elastic potential. The higher-derivative curvature term, with coefficient $K$ will play an important role below; ellipses denote further higher-derivative terms. 
Expanding the potential $V_{\rm el}$ up to quadratic order leads to a Gaussian theory of linearly dispersing longitudinal and transverse phonons
\begin{equation}\label{eq_L2}
\begin{split}
\mathcal L_{(2)} = \frac{\chi}2 \Bigl[(\partial_t u_i)^2 - c_T^2 (\partial_i u_j)^2 & -  (c_L^2 - c_T^2)(\partial_i u_i)^2\\
	& + K (\nabla^2 u_i)^2 + \cdots \Bigr]\, ,
\end{split}
\end{equation}
with velocities $c_L$ and $c_T$ related to the shear and bulk moduli as $\mu = \chi c_T^2$ and $B = \chi(c_L^2 - \frac{2(d-1)}{d} c_T^2)$.
Treated as a systematic EFT, Eq.~\eqref{eq_L} is a translational and rotational symmetry-preserving (in reference and target spaces) expansion, both in gradients and powers of the phonon fields. Let us determine how important these latter nonlinearities, i.e.,  phonon interactions,  are. For excitations with dispersion $\omega\sim q^z$ ($z=1$ for conventional phonons above), at the Gaussian fixed point the kinetic action  $S_{(2)} \sim \int dt d^d x\, (\partial_t u_i)^2 \sim 1$, thus sets the amplitude of phonon fluctuations to scale as $u_i^2 \sim q^{d-z}$. Hence, the relative importance of nonlinearities ${\cal L}_{(n>2)}\sim (\partial_i u_j)^n$ is given by
\begin{equation}\label{eq_scaling}
\frac{\mathcal L_{(n>2)}}{\mathcal L_{(2)}}
	\sim \frac{(\partial_i u_j)^n}{(\partial_t u)^2} 
	\sim q^{\frac12(n-2)(d-z)+n-2z}\, .
\end{equation}
For conventional phonons with $z=1$ this ratio reduces to $q^{(n-2)(d+1)/2}$, recovering the well-known irrelevance of nonlinearities ${\cal L}_{n>2}$ in any dimension $d$.

However, here we reexamine this conclusion for a phonon velocity tuned to zero, so that the Gaussian action (\ref{eq_L2})
is characterized by $z=2$ dynamics. A similar problem was studied for a {\em scalar} Nambu-Goldstone field $\theta$  in \cite{Yang:2004ferr,Vishwanath:2003yjl,Kozii:2017ferr,  Ardonne:2003wa,Fradkin:2003vaq}. In this context, the leading nonlinearity allowed by rotation invariance and charge conjugation symmetry is quartic, $[(\partial_i\theta)^2]^2$
	\footnote{In the absence of charge conjugation symmetry $\theta \to -\theta$, an additional marginal term is allowed in $d=2$: $\dot \theta (\nabla \theta)^2$  \cite{Kozii:2017ferr}. }.
Setting $n=4$ and $z=2$ in Eq.~\eqref{eq_scaling}, such 
a quartic nonlinearity scales as $q^{d-2}$, becoming relevant in spatial dimensions $d<2$. It is marginally irrelevant in $d=2$, thereby allowing for a controlled $\epsilon = 2-d$ expansion toward a nontrivial $d=1$ critical point. This fixed point was originally proposed to describe the ferromagnetic instability of a 1D Luttinger liquid \cite{Yang:2004ferr, Kozii:2017ferr} %
	\footnote{However, the coupling of the Lifshitz spin modes to the gapless charge degrees of freedom may make the transition first order in this context \cite{Zedler:2005ferr}.}. 
The same universality class is expected to describe the IR of the Ising CFT and compact scalar CFT deformed by the Lorentz symmetry-breaking operator $\sigma\partial_x \phi$ \cite{Cole_2019}. 
The Lifshitz scaling with nontrivial fixed point allows for a loophole to the Coleman-Hohenberg-Mermin-Wagner theorem in $d=1$ \cite{NelsonPeliti, AronovitzLubensky, DG,LeDoussal:1992xh}.
The marginal universality in $d=2$ was first studied in a 3D classical smectic model in Refs.~\cite{PhysRevLett.47.856,PhysRevA.26.2196}. In the quantum context, it describes the deconfined quantum critical point between two different valence bond solid phases \cite{Vishwanath:2003yjl,Ardonne:2003wa,Fradkin:2003vaq}%
	\footnote{In the context of the deconfined quantum critical point, the superfluid phase $\theta$ is the dual photon. However, the magnetic $U(1)$ symmetry protecting it only emerges at the Lifshitz QCP (where monopoles are suppressed), so that the neighboring phases are both gapped.}
and quantum smectics as realized in FFLO superconductors \cite{ffloLR}. It also arises in the context of stochastic quantization \cite{Dijkgraaf:2009gr}.
We refer to this universality class as the {\em scalar Lifshitz QCP}. 

Inspired by these studies, we now consider a similar $z=2$ limit in the EFT of quantum elasticity. In isotropic solids, positivity  of the bulk modulus $B$ (required by stability) implies that transverse phonons are always slower than longitudinal phonons. We thus consider a quantum solid with $c_T\to 0$. The fact that the phonon vector field $u_i$  carries a spatial index allows for {\em cubic} nonlinearities even in isotropic solids, which are more relevant than the quartic interactions in the scalar Lifshitz QCP.
For $n=3$ and $z=2$,  Eq.~\eqref{eq_scaling} leads to $q^{(d-4)/2}$, so that cubic nonlinearities are relevant in $d<4$. Thus, floppy quantum solids are strongly interacting, even in $d=3$. We dub this new universality class as the {\em crystal Lifshitz QCP}.

The RG fate of $d=4$ marginal cubic interactions is more subtle than that of quartic ones. We recall that in a Euclidean time path integral, it is clear that a single positive {\em quartic} coupling tends to decrease at 1-loop order after integrating out a Wilsonian shell, giving $\delta \lambda = - c \lambda^2$, with $c>0$. This leads to the well-known marginal irrelevance of interactions in $\phi^4$ theory, as well as in the scalar Lifshitz QCP discussed above. In contrast, a similar argument suggests that {\em cubic} couplings receive instead a positive self-correction, $\delta g = c g^3$. However, cubic interactions also renormalize the propagator at 1-loop order, which feeds back into $\delta g$, leading to a competition. In the case of the scalar $\phi^3$ theory, vertex renormalization wins and the cubic coupling is marginally relevant in the infrared, precluding a fixed point below its upper-critical dimension \cite{srednicki2007quantum}. In stark contrast, for the crystal Lifshitz universality class, we will demonstrate that the negative contribution from the renormalization of the propagator wins over the positive vertex correction, and the coupling is marginally {\em irrelevant}. This thus allows us to construct a perturbatively accessible critical point in $d=4-\epsilon$ spatial dimensions.

\section{Summary of critical results}

Before turning to a technical analysis, we summarize our key predictions for a quantum critical solid. Its strongly coupled transverse phonons exhibit a universal scaling behavior entering in a variety of observables. For example, they dominate the dynamic structure factor at low frequencies, for wavevectors near a nonzero reciprocal lattice vector ($q\ll G$)
\begin{equation}
S(\omega,G+q) \sim \frac{1}{q^{2z}} \hat f_s(\omega/q^z)\, ,
\end{equation}
with $\hat f_s(x)$ a universal dimensionless scaling function, with $\hat f_s(x \ll 1)\rightarrow {\text {const}}$, and $\hat f_s(x \gg 1)\rightarrow 1/x^2$. They also dominate the specific heat,
\begin{equation}
C_V \sim T^{d/z}\, .
\end{equation}

We also predict that quantum critical solids (characterized by a vanishing shear modulus), exhibit a universal  non-Hookean i.e., {\em nonlinear} elastic response of strain $\epsilon = \langle \partial_x u_y\rangle$ to shear stress $\sigma$:
\begin{equation}
\epsilon
	\sim \sigma^{1/\delta}\, , \qquad
	\delta  = \frac{d+3z - 2}{d + 2 - z} > 1\,.
\end{equation}
Linear response is recovered for nonzero temperature $T$ and/or detuning away from criticality $c_T^2 = p-p_c$, with a singular shear modulus $\mu = d\sigma / d\epsilon|_{\epsilon = 0}$
\begin{equation}
\mu \sim T^{2(z-1)/z}\, , \qquad
\mu \sim (c_T^2)^{2\nu(z-1)}\, .
\end{equation}
It also shows a universal dynamic form
\begin{equation}
\mu(\omega,T)
    \sim \omega^{2(z-1)/z} \hat f_\mu(\omega/T)\, ,
\end{equation}
and corresponding shear viscosity $\eta(\omega,T) = \frac{\mu(\omega,T)}{-i\omega}$, with $\hat f_\mu(x)$ another universal scaling function. 

The expressions above can be summarized by a scaling form for strain
\begin{equation}
\epsilon(\sigma, c_T^2, T, \omega,q)
	 = \sigma^{1/\delta} \mathcal A \left(\frac{c_T^{2\nu z}}{T}, \frac{\sigma^{2z}}{T^{d-2+3z}}, \frac{\omega}{q^z},\frac{\omega}{T}\right)\, .
\end{equation}
The exponents $z$ and $\nu$ are the dynamic and correlation length exponents, which are universal properties of the 3+1d strongly-coupled Lifshitz QCP (as is the entire scaling function $\mathcal A$). We obtain these exponents below within an $\epsilon$-expansion, see \eqref{eq_z} and \eqref{eq_Delta_O}. 

As we detail below, this universal phenomenology applies inside the Ginzburg region, for $c_T > c_T^c$. However, for $c_T < c_T^c$, we predict that the cubic nonlinearity drives the solid through a first-order structural phase transition, with the distorted state dependent on the details of the soft crystal.

\section{Lifshitz Solid QCP in $d=4-\epsilon$}

We now turn to the analysis that leads to the above results. To this end, it is convenient to separate phonon fields into orthogonal transverse and longitudinal components, $u^i = u_i^T + u_i^L$, with $\partial_i u_i^T = 0$. Because a floppy solid is characterized by $c_T\rightarrow 0$, we focus on the critical {\em transverse} phonons, $u_i^T(t,{\bf x})$, controlled by an effective Lagrangian (see App.~\ref{app_EFT}\ for details)
\begin{equation}\label{eq_S_4d}
{\cal L} = \frac12 u_i^T\!\left(\partial_t^2 - c_T^2 \nabla^2 + K \nabla^4\right)\!u_i^T
	+ g \partial_i u^T_j \partial_i u^T_k \partial_j u^T_k ,
\end{equation}
that emerges from (\ref{eq_L}), keeping the leading interactions. Higher-order nonlinearities are irrelevant in the RG sense, and dropping total derivative terms such as $\partial_i u^T_j \partial_j u^T_k \partial_k u^T_i$ leaves us with a single cubic coupling. Note that one cannot strictly ignore the longitudinal modes, since they are gapless; however, one can show perturbatively that they do not renormalize the transverse sector.

We are interested in the IR behavior of this theory in the formal critical limit of $c_T\to 0$ (but see below). First note that the quadratic time-derivative term does not get renormalized, and we therefore set its coefficient to $\chi=1$. Indeed, any diagrammatic correction to the propagator coming from the above vertex
$g$ involves at least two spatial gradients on the external legs, so that only $(\partial_i u_j^T)^2$ and higher-derivative terms get generated under RG. Thus, the momentum susceptibility remains finite at the QCP.

The IR fate of model \eqref{eq_S_4d} is thus determined by the renormalization of $g$ and $K$, or more precisely by the dimensionless coupling $\hat g =\frac1{\sqrt{8\pi^2}}\Lambda^{(d-4)/2} g/K^{5/4}$ (the factor of $8\pi^2$ is introduced for convenience, to absorb the volume factors arising in loops).  Integrating out a shell of momenta $\Lambda/b < p < \Lambda$ (with $b>1$), we find the renormalization of the cubic coupling $g$ is (see App.~\ref{app_RG_d4} for details)
\begin{equation}
\frac{d \log g}{d \log b} \simeq \!\!
\vcenter{\hbox{
  \begin{tikzpicture}[scale=0.6, line width=0.8pt]
    \coordinate (v1) at (90:0.6);
    \coordinate (v2) at (210:0.6);
    \coordinate (v3) at (330:0.6);
    
    \draw (v1) -- (v2) -- (v3) -- cycle;
    
    \draw (v1) -- (90:1.2);
    \draw (v2) -- (210:1.2);
    \draw (v3) -- (330:1.2);
  \end{tikzpicture}
}}
\!\! = \gamma_g \hat g^2 \, , \quad \ \gamma_g = \frac{1}{64}, 
\end{equation}
showing that the cubic coupling tends to grow, as discussed in the previous section. However, the propagator also gets renormalized already at one-loop order:
\begin{equation}\label{eq_RG_dK}
\frac{d\log K}{d\log b} \simeq \!\!
\vcenter{\hbox{
  \begin{tikzpicture}[scale=0.6, line width=0.8pt]
    \coordinate (v1) at (-0.6, 0);
    \coordinate (v2) at (0.6, 0);

    \draw (v1) -- (-1.2, 0);
    \draw (v2) -- (1.2, 0);

    \draw (v1) to[out=60, in=120] (v2);
    \draw (v1) to[out=-60, in=-120] (v2);
  \end{tikzpicture}
}}\!\!
= \gamma_K \hat g^2\, , \quad \  \gamma_K = \frac{43}{192} ,
\end{equation}
giving a {\em negative} correction to $\hat g$. 
The resulting $\beta$-function 
thus features a competition between these two effects:
\begin{equation}
\beta_{\hat g} \equiv \frac{d \hat g}{d \log \frac1b}
	\simeq \left(\frac54 \gamma_K - \gamma_g\right) \hat g^3 = \frac{203}{768} \hat g^3\, .
\end{equation}
The renormalization of the kinetic term dominates with a significant margin, overturning the tendency of the cubic coupling to grow at long length scales %
	\footnote{Qualitatively, this outcome opposite to $\phi^3$ in $d=6$ arises because of the gradients in the interaction: loop integrals over momentum now involve spherical averages of products of unit vectors. Vertex renormalization is smaller because it involves more vectors.}. 
The cubic nonlinearity is thus marginally irrelevant in $d=4$, leading to a perturbatively accessible critical point in $d=4-\epsilon$, in striking contrast with scalar $\phi^3$-theory.

Including the length and time rescaling (engineering dimension of $u^T_i$ at the Gaussian critical point) gives a $\beta$-function in $d=4-\epsilon$
\begin{equation}
\beta_{\hat g}
	\simeq -\frac{\epsilon}2 \hat g + \left(\frac54 \gamma_K - \gamma_g\right) \hat g^3 
\end{equation}
that vanishes at the fixed point coupling $\hat g_*^2 \simeq \frac{2\epsilon}{5\gamma_K - 4 \gamma_g} = \frac{384}{203}\epsilon$. The dynamic critical exponent $z$ of this Lifshitz-solid QCP arises from the renormalization of $K$, which changes the scaling of time with respect to space. The dispersion relation of transverse phonons now becomes $\omega = \sqrt{K+\delta K}q^2$, with $\delta K$ given by \eqref{eq_RG_dK}, so that
\begin{equation}
\omega \simeq \sqrt{K}\left(1 - \frac12 \gamma_K \hat g^2 \log q\right)q^2\, .
\end{equation}
Evaluating this at the critical coupling $\hat g_*$ and identifying with $\omega = \sqrt{K} q^{z}\simeq\sqrt{K} q^{2}(1+(z-2)\log q)$ leads to
\begin{equation}\label{eq_z}
z 
	\simeq 2 - \frac{\gamma_K}{5\gamma_K - 4\gamma_g} \epsilon
	= 2 - \frac{43}{203} \epsilon\, .
\end{equation}
It is interesting to contrast this with the {\em scalar} Lifshitz QCP, where $z$ only differs from $2$ in $d < 2$ and at two-loops \cite{Sachdev:1996zero,Yang:2004ferr,Kozii:2017ferr}. In $d=1$, the resulting deviation $\delta z / z$ is thus expected to be very small, consistent with current numerics \cite{Cole_2019}. Extrapolating our one-loop results to $d=3$ suggests that floppy solids have an appreciable anomalous dynamic critical exponent, $\delta z /z \approx 0.1$.

One can similarly obtain the correlation length critical exponent $\nu$, or, equivalently, the dimension of the relevant operator $\mathcal O=(\partial_i u_j^T)^2$ corresponding to the tuning parameter $c_T^2$. We compute this dimension in App.~\ref{app_RG_d4} and quote the result here:
\begin{equation}\label{eq_Delta_O}
\Delta_{\mathcal O}
	\equiv d+ z - \frac{1}{\nu}
	\simeq 4- \frac{256}{203} \epsilon\,\  \Rightarrow\   \nu = \frac12(1-\frac{5}{203}\epsilon)\;
\end{equation}
The correlation length critical exponent $\nu$ is very close to its mean field value, due to a near cancellation between the renormalization of the parameter $c_T^2$ and  that of $z$.

\section{Ginzburg criterion}

As anticipated in the introduction, the cubic coupling responsible for this new quantum critical point also has the unsavory consequence of generically driving the transition first-order. Indeed, tuning $c_T\to 0$ to reach the critical point leads to an unbounded potential $\sim g (\partial u)^3$. However, if the {\em quartic} coupling $\lambda$ is large enough to stabilize the field $u$ at a small enough value, fluctuations of $u$ may be large enough for the critical data of the crystal Lifshitz QCP to be visible even when slightly off-criticality. This is the Ginzburg criterion, which we spell out below.

Let us start from the Gaussian fixed point with $c_T, g =0$. Turning on the relevant deformation $g$ defines a UV momentum scale $\Lambda_{\rm UV} = g^{2/\epsilon}$ below which the system becomes strongly interacting (we work in units of time where $K=1$). Even if the tuning parameter $c_T$ is not exactly zero, there is a parametric window of momenta controlled by the above nontrivial crystal Lifshitz universality, 
as long as the associated IR scale $\Lambda_{\rm IR} = c_T$ is parametrically smaller than $\Lambda_{\rm UV}$:
\begin{equation}\label{eq_Ginzburg}
c_T \ll g^{2/\epsilon}\, .
\end{equation}
A nonzero $c_T$ can avoid the first-order transition at $c_T^c$:
for a potential $c_T^2 (\partial u)^2 + g (\partial u)^3 + \lambda (\partial u)^4$, the instability is avoided if $c_T^2 \gtrsim (c_T^c)^2\equiv g^2 / \lambda$. Combining with \eqref{eq_Ginzburg} leads to the condition
\begin{equation}
\frac{g^2}{\lambda} \ll g^{4/\epsilon}
\end{equation}
for the quantum critical point to be visible in an intermediate range of momenta $\Lambda_{\rm IR} \lesssim q \lesssim \Lambda_{\rm UV}$. This could be achieved with a moderate amount of fine-tuning of the quartic coupling $\lambda$.

\section{Floppy solids in $d=2$}

Our analysis so far has revealed the existence of a perturbative crystal Lifshitz QCP in $d=4-\epsilon$, which may control the low-energy dynamics of floppy solids in $d=3$ spatial dimension. However, the behavior of floppy solids in $d=2$ is entirely different: the cubic coupling in Eq.~\eqref{eq_S_4d}, which was responsible for the RG flow discussed above, vanishes identically in two dimensions (this is simplest to see using the parametrization \eqref{eq_uT_d2}). The leading nonlinearities involving transverse phonons are then quartic. Eq.~\eqref{eq_scaling} shows that quartic nonlinearities have upper critical dimension $d = 2$. Transverse phonon interactions are thus again marginal! We study their IR fate below.

The Lagrangian for transverse phonons in $d=2$ is simplest using the parametrization
\begin{equation}\label{eq_uT_d2}
u_i^T = \epsilon_{ij}\partial_j \phi\, ,
\end{equation}
and takes the form
\begin{equation}\label{eq_S_2d}
\begin{split}
\mathcal L
	&=  \frac{\chi}2 \nabla\phi \left(-\partial_\tau^2 - c_T^2 \nabla^2 + K \nabla^4\right) \nabla\phi \\
	&\!\!\!+ \lambda_1 (\nabla^2 \phi)^4 + \lambda_2 [(\partial_i \partial_j \phi)^2]^2
	+ \lambda_3 (\partial_i \partial_j \phi)^2 (\nabla^2 \phi)^2.
\end{split}
\end{equation}
App.~\ref{app_RG_d2} shows how the three quartic couplings $\lambda_{1,2,3}$ relate to more familiar nonlinearities in the original strain field $u_{ij}$ of elasticity theory. Quantum critical solids in $d=2$ thus have a rich landscape of possible IR outcomes, depending on the region of the parameter space $(\lambda_1,\lambda_2,\lambda_3)$. The one-loop $\beta$-functions
\begin{equation}\label{eq:betafn_d2_1}
\beta_i = C_{ijk} \lambda_j \lambda_k\, , 
\end{equation}
computed in App.~\ref{app_RG_d2}, lead to: (A) marginally irrelevant regions that flow back to the Gaussian fixed point $\lambda_i = 0$, and (B) regions flowing to a unstable parameters where the Hamiltonian \eqref{eq_S_2d} becomes unbounded below. Two-dimensional floppy solids with parameters in region (A) are weakly coupled in the IR, with logarithmic corrections that can be resummed as has been done for the scalar Lifshitz universality class in $d=2$ \cite{PhysRevLett.47.856,PhysRevA.26.2196,Vishwanath:2003yjl,Ardonne:2003wa,Fradkin:2003vaq,Yang:2004ferr,criticalMatterLR}. If the parameters are in region (B), the floppy solid most likely exhibits a fluctuation-driven first-order structural transition. One other interesting feature of the $d=2$ crystal Lifshitz QCP is the presence of vertex operators $e^{i \vb G \cdot \vb u}$ carrying momentum $\vb G$ on the reciprocal lattice, and with scaling dimension set by the dimensionless ratio $G^2/(\chi\sqrt{K})$ playing a similar role to the Luttinger parameter of Luttinger liquids. These are further discussed in App.~\ref{app_obs}.

\section{Extensions and Applications}

Above we discussed a discovery of a strongly-coupled Lifshitz critical universality class, even in the simplest {\em isotropic} quantum solids. However, crystals are characterized  by a variety of point group symmetries, and thus allow for numerous generalizations with different critical-solid universality classes. One such generalization is to consider phonon velocities tuned to vanish only for momenta $\vb q$ on a submanifold of dimension $m < d$ ($m$-Lishitz models \cite{criticalMatterLR}). In fact, with multiple phonons, different phonons may exhibit different dimensions $m$, admitting a rich set of possibilities. In this case, interactions are typically irrelevant, with the IR then captured by a crystal-Lifshitz Gaussian fixed point, as was discussed long ago \cite{Komatsu1951,Lifshitz1952,PhysRevLett.115.025703}.

However, there are scenarios where the quantum dynamics can flow to a strongly coupled critical point similar to the isotropic ones studied above. For a dispersion relation $\omega^2 \sim  \vb \sum_{i=1}^m q_{i}^4 + \sum_{i=m+1}^d q_i^2$, a simple extension of our scaling argument above shows that interactions $(\partial u)^{2+n}$ are marginal for $2d = m + \frac{4}{n}$. Quartic ($n=2$) interactions are thus always irrelevant if $m<d$ (unlike the $d=m=2$ situation that we studied above). Cubic ($n=1$) interactions however are not necessarily irrelevant: they are relevant for $d=2,\,m=1$ and marginal for $d=3,\,m=2$. This last situation arises in layered Van der Waals materials, which are close to realizing $m = 2$ Lifshitz phonons without the need for fine tuning. A flow to a nontrivial critical point and the associated anomalous critical elasticity will be realized provided the point group symmetry allows for a cubic elastic nonlinearity,
\begin{equation}
T^{ijk} \partial_i u^z\partial_j u^z \partial_k u^z\, , 
\end{equation}
with $T^{ijk}$ a rank-3 tensor with indices in the $ab$-plane. Such a term is allowed, for example, in rhombohedral graphite, 3R-stacked TMDs, and 1T-TiS${}_2$. Other potential realizations of quantum critical solids include materials near a quantum structural phase transition, quantum phase transitions with an order parameter coupling linearly to strain \cite{PhysRevLett.115.025703}, or Wigner crystals  near the quantum melting transition (see, e.g., \cite{Grover:2025rsv}).
We leave a detailed exploration of such anisotropic elastic nonlinearities and other generalizations for future studies.

Our construction may also have more formal applications. Doped conformal field theories are an interesting platform to generate compressible phases of matter with universal transport parameters (or EFT Wilsonian coefficients) \cite{Son:2002zn,Sachdev:2012tj,Hellerman:2015nra,Delacretaz:2025ifh,Divic:2025yvf}. When a doped CFT becomes a superfluid, scale invariance is known to fix its Goldstone mode velocity to be $1/\sqrt{d}$ in units of the CFT velocity. Instead, conformal solids maintain a tuning parameter, that can give access to the floppy limit \cite{Esposito:2017qpj}. It would be interesting to find examples of doped CFTs that realize this physics.

Finally, the floppy limit $c_T\to 0$ of solids has also been considered from a very different perspective in Refs.~\cite{Endlich:2010hf,Goldberger:2025mgb} to try to makes sense of hydrodynamics as a $T=0$ quantum field theory. In that model, the entire Hamiltonian for transverse modes is taken to zero; nevertheless, see \cite{Dersy:2022kjd,Cuomo:2024ekf} for possible Lifshitz-like physics in that context.

\vspace{5mm}
\section*{Acknowledgments}
We thank Gabriel Cuomo, Ilya Esterlis, Eduardo Fradkin, Tarun Grover, Austin Joyce, Steve Kivelson, John McGreevy, Riccardo Penco, Jay Deep Sau, and Ruben Verresen for helpful discussions. LR thanks John Toner and  Pierre Le Doussal for discussions and collaborations on related topics. 
LD is supported by a NSF CAREER award (DMR-2441227) and a Sloan Fellowship. LR is supported by the Simons Investigator award from the Simons Foundation. S.D.C. is supported by ``Exotic High Energy Phenomenology'' (X-HEP), a project funded by the European Union -- Grant Agreement n.~101039756 (PI: J.~Elias~Mir\'o). Views and opinions expressed are however those of the author(s) only and do not necessarily reflect those of the European Union or the ERC Executive Agency (ERCEA). Neither the European Union nor the granting authority can be held responsible for them. AI models (Claude, Gemini, ChatGPT) and Mathematica were used to check the results of the RG calculations.



%

\clearpage
\onecolumngrid

\appendix

\makeatletter

\renewcommand{\thesubsection}{\thesection.\arabic{subsection}}

\renewcommand{\subsection}{%
  \@startsection
    {subsection}%
    {2}%
    {\z@}%
    {.8cm \@plus1ex \@minus .2ex}%
    {.5cm}%
    {\normalfont\small\itshape\centering}%
}

\makeatother
\section{Review of Elasticity Theory}\label{app_EFT}

Let us label the material points of the undistorted solid
by $\vb x\in\mathbb{R}^d$ and let $R^i(\vb x,t)=\vb x^i+u^i(\vb x,t)$ be the position of
the atom labelled $\vb x$, with $u^i$ the dynamical field. Two independent groups
act: rigid motions of physical space, $R^i\to O^{ij}R^j+a^i$, under which $u^i$
shifts by $(O-\mathbb{I})^{ij}\vb x^j$ and relabelings of the reference configuration,
$\vb x^i\to\tilde O^{ij}\vb x^j+b^i$.  At lowest
order in derivatives, invariance under the rigid motion of the physical space constrains the  building blocks to be the induced metric
and the velocity bilinear,
\begin{equation}
  g_{ij}\equiv\partial_iR^k\partial_jR^k=\delta_{ij}+2u_{ij},
  \qquad \dot R^i\dot R^j=\dot u^i\dot u^j,
\end{equation}
with
\begin{equation}
  u_{ij}=\tfrac12\big(\partial_iu_j+\partial_ju_i
  +\partial_iu_k\partial_ju_k\big)\equiv\varepsilon_{ij}+\tfrac12M_{ij}
  \label{eq:strain}
\end{equation}
the Green-Lagrange strain, where
$\varepsilon_{ij}=\tfrac12(\partial_iu_j+\partial_ju_i)$ and
$M_{ij}=\partial_iu_k\partial_ju_k$. Relabellings of the reference 
configuration ensures that the terms appearing in the lagrangian are rotationally
invariant with respect to the reference frame. Hence
$\mathcal{L}_E=\chi\tfrac12\,\dot u_i\dot u_i+F$ with
$F=F_{\rm el}(u_{ij})+F_\nabla(\partial_ku_{ij}, \dot u_i)$. The elastic free-energy density $F_{\rm el}$ admits a systematic expansion in products of rotational invariants constructed from the strain tensor, such as $(\tr u)^m$ and $\tr u^n$. We impose parity invariance, excluding operators containing an odd number of Levi-Civita tensors. The first few terms in expansion of $F_{\rm el}$ is given by,
\begin{align}
  F_{\rm el}=\;&\sigma\tr u+\tfrac{\lambda}{2}(\tr u)^2+\mu\tr u^2+\nu_1(\tr u)^3+\nu_2(\tr u)(\tr u^2)+\nu_3\tr u^3+ O(u^4).
  \label{eq:Fel}
\end{align}
The first few Wilsonian coefficients entering the EFT of elasticity above have names: $\lambda,\mu$ are the Lam\'e coefficients ($\mu$ is the shear modulus), and $\sigma$ is sometimes referred to as a residual pre-stress; in terms of these coefficients, the bulk modulus is $B=\left( \lambda + 2 \frac{2 \mu}{d}+\left( \frac{2}{d}-1\right)\sigma\right)$. $\nu_{1,2,3}$ are higher-order
elastic constants. 

The gradient contribution $F_\nabla$ contains rotationally invariant contractions of $u_{ij}$, $\dot u_{i}$ and their spatial derivatives, such as $\partial_k u_{ij}, \partial_k\partial_l u_{ij}, \partial_k \dot{u}_i$ and so on. The leading spatial gradient corrections to $F_\nabla$ are quadratic in
$\partial_ku_{ij}$. Several invariants of this type exist, but at quadratic order
in $u$ each is a rotational scalar of degree four in momentum, hence a
combination of the two structures $q^4\delta_{ij}$ and $q^2q_iq_j$.  Decomposing $u^i=u_i^L+u_i^T$
with $\nabla\!\cdot\!u^T=0$, this implies that the leading terms can be constructed out of
$(\partial_i\nabla\!\cdot\!u)^2$ and $(\nabla^2u_j)^2$. The former
only gives a subleading (in $q^2$) correction to the dispersion of longitudinal modes and we will not consider it further in this analysis. The latter, whose coefficient we call $\chi K$, becomes the leading
transverse restoring force once $c_T$ is tuned to zero. Finally, we omit additional non-linear operators involving time derivatives, anticipating Lifshitz $z > 1$ scaling, since they would be less relevant than spatial derivatives.

\subsection{Gaussian theory and the floppy limit}

Decomposing $u_i=u_i^L+u_i^T$
with $\nabla\!\cdot\!u^T=0$ and collecting the terms of \eqref{eq:Fel} quadratic
in the fields, one finds the gaussian Euclidean action,
\begin{align}
  S^E_{(2)}= \chi\int\dd\tau\,\dd^dx\Big[
  &\tfrac{1}{2}\dot{u}_i^2
  +\tfrac{c_L^2}{2}(\nabla\!\cdot\!{u}_i^L)^2+\tfrac{c_T^2}{2}(\partial_iu_j^{T})^2
  +\tfrac{K}{2}(\nabla^2u_j^{T})^2\Big],
  \label{eq:S2}
\end{align}
with
\begin{equation}
  \chi c_L^2=\left(\lambda+2\mu+\sigma\right),
  \qquad
  \chi c_T^2=\left(\mu+\sigma\right) .
  \label{eq:speeds}
\end{equation}
The floppy limit we consider is $c_T^2\to 0$, where
\begin{equation}
  \omega_T(q)=\sqrt{K}\;q^2,
  \qquad
  \omega_L(q)=c_Lq ,
  \label{eq:disp}
\end{equation}
so that the soft sector has dynamical exponent $z=2$. For small nonzero $c_T$ the quadratic dispersion crosses over to linear below $q_*=c_T / \sqrt{K}$. The phonon propagators are 

\begin{equation}\label{eq_u_propagator}
\langle u^T_i u^T_j\rangle(p)
	=  \frac{P_{ij}(p)}{\chi\left(p_0^2 + c_T^2 (\vb p\cdot \vb p)  + K (\vb p\cdot \vb p)^2\right)}\, , \quad \langle u^L_i u^L_j\rangle(p)
	= \frac{\delta_{ij}-P_{ij}(p)}{\chi\left(p_0^2 + c_L^2 (\vb p\cdot \vb p)\right)}\,,\quad 
	P_{ij}(p) \equiv \delta_{ij} - \frac{p_i p_j}{\vb p\cdot \vb p}\, .
\end{equation}
where $P_{ij}(p)$ denotes the transverse projector, and $p_0$ is the Euclidean frequency.

\subsection{The cubic vertex and $d=4$}\label{eq:cubic_d4_app}
As discussed in the main text, the cubic vertex constructed out of $u^T_i$ is marginal in $d=4$ when the transverse phonon has a $z=2$ propagator  \eqref{eq:disp}. Instead, the longitudinal phonon stays weakly coupled in that dimensionality. Using the method of regions, it is possible to show that longitudinal vertices are nevertheless renormalized (composite longitudinal operators acquire anomalous dimensions) due to their coupling to the strongly coupled sector transverse sector. However, the longitudinal sector does not affect in turn the transverse sector, and is merely a spectator to the RG flow. This justifies focusing on the dynamics of transverse phonons alone: we therefore set $u_i^L=0$ and work with a purely transverse field,
$\nabla\!\cdot\!u^T=0$.

We expand \eqref{eq:Fel} in powers of the field using
$u_{ij}=\varepsilon_{ij}+\tfrac12M_{ij}$. For a transverse configuration
$\tr\varepsilon=\nabla\!\cdot\!{u}_T=0$, so $\tr u=\tfrac12\tr M$ is already
quadratic and the invariants $\lambda$, $\nu_1$ and $\nu_2$ contribute
only at quartic order and beyond. The cubic terms come from $\mu\tr u^2$ and
$\nu_3\tr u^3$,
\begin{equation}
  F^{(3)}=\mu\,\varepsilon_{ij}M_{ij}+\nu_3\tr\varepsilon^3 .
  \label{eq:cubic}
\end{equation}
The two structures are not independent. Expanding $\varepsilon_{ij}$ and using
the symmetry of $M_{ij}$,
\begin{align}
  \varepsilon_{ij}M_{ij}&=\partial_iu_j\,\partial_iu_k\,\partial_ju_k,\qquad
  \tr\varepsilon^3=\tfrac14\,\partial_iu_j\,\partial_ju_k\,\partial_ku_i
  +\tfrac34\,\varepsilon_{ij}M_{ij},
  \label{eq:eps3}
\end{align}
The eight terms of the second expansion can be organised into two types by cyclicity and transposition. The first structure of $  \tr\varepsilon^3$ 
integrates to zero for a transverse field
\begin{equation}
  \int\partial_iu_j\,\partial_ju_k\,\partial_ku_i
  =-\!\int\! u_j\big[\partial_i\partial_ju_k\,\partial_ku_i
  +\partial_ju_k\,\partial_k(\partial_iu_i)\big],
\end{equation}
where the second term vanishes by transversality, and since
$\partial_j(\partial_iu_k\partial_ku_i)
=2\,\partial_i\partial_ju_k\,\partial_ku_i$ a second integration by parts leaves
$\tfrac12\int(\partial_ju_j)\,\partial_iu_k\partial_ku_i=0$. Hence the transverse
cubic vertex is unique,
\begin{equation}
  F^{(3)}=\chi g\,\partial_iu_j^{T}\,\partial_iu^T_{k}\,\partial_ju^T_{k},
  \qquad
  \chi g=\mu+\tfrac34\nu_3 ,
  \label{eq:g3}
\end{equation}
and the critical theory near four dimensions is
\begin{align}
  S_{d=4}=\chi\int\dd\tau\,\dd^dx\Big[
  &\tfrac{1}{2}\dot{u}_i^2
  +\tfrac{  c_L^2}{2}(\nabla\!\cdot\!{u}_i^L)^2+\tfrac{c_T^2}{2}(\partial_iu_j^{T})^2+\tfrac{  K}{2}(\nabla^2u_j^{T})^2+g\,\partial_iu_j^{T}\partial_iu_k^{T}\partial_ju_k^{T}\Big],
  \label{eq:Sd4}
\end{align}
with $\chi  c_T^2$ the single relevant coupling. In $d=4$, this reproduces the starting point of the analysis in  \eqref{eq_S_4d} of the main text. 

\subsection{Quartic vertices and $d=2$}\label{app_EFT_IC} 

We now turn to $d=2$, where transversality is solved by
$u_i^{T}=\epsilon_{ij}\partial^j\phi$ and $\epsilon_{ij}$ is the
two dimensional Levi-Civita tensor. Let us define,
\begin{equation}
 H_{ij}=\partial_i\partial_j\phi,\qquad h_1\equiv(\tr H)^2=(\nabla^2\phi)^2,
  \qquad
  h_2\equiv\tr H^2=(\partial_i\partial_j\phi)^2.
\end{equation}

Since  $\varepsilon_{ij}$ is a
symmetric traceless $2\times2$ matrix both structures in \eqref{eq:cubic} vanish identically. 

Therefore, the leading transverse interactions are quartic, and with respect to the $z=2$ transverse
propagator they are marginal in two spatial dimensions.

In $d=2$, the Cayley-Hamilton theorem 
$u^2=(\tr u)\,u-(\det u)\mathbb{I}$, expresses every trace beyond $\tr u^2$ in
terms of $\tr u$ and $\tr u^2$,
\begin{align}
  \tr u^3&=\tfrac32\,\tr u\,\tr u^2-\tfrac12(\tr u)^3, \qquad
  \tr u^4=\tfrac12(\tr u^2)^2+(\tr u)^2\,\tr u^2-\tfrac12(\tr u)^4 .
  \label{eq:CH}
\end{align}
Continuing \eqref{eq:Fel} to quartic order, the invariants there are
$(\tr u)^4$, $(\tr u)^2\tr u^2$, $(\tr u^2)^2$, $\tr u\,\tr u^3$ and $\tr u^4$,
with couplings $w_1,\dots,w_5$. Eliminating $\tr u^3$ and $\tr u^4$ by
\eqref{eq:CH}, and absorbing $\nu_3$ into $\nu_{1,2}$ and $w_{4,5}$ into
$w_{1,2,3}$, ~\eqref{eq:Fel} becomes
\begin{align}
  F_{\rm el}=\;&\tfrac{\lambda}{2}(\tr u)^2+\mu\,\tr u^2
  +\nu_1(\tr u)^3+\nu_2\,\tr u\,\tr u^2 \nonumber\\
  &+w_1(\tr u)^4+w_2(\tr u)^2\tr u^2+w_3(\tr u^2)^2+O(u^5)
  \label{eq:Fd2}
\end{align}
where the signs of the purely quartic terms have been chosen so that positive values of the coupling lead to a potential bounded from below. 
Only $\lambda,\mu,\nu_2$ and $w_3$ survive below. Evaluating the two invariants
on a transverse configuration with \eqref{eq:strain},
\begin{align}
  \tr u&=\tfrac12h_2, \qquad
  \tr u^2 =\Big(h_2-\tfrac12h_1\Big)
  +\tfrac14\Big(\tfrac12h_2^2+h_1h_2-\tfrac12h_1^2\Big),
  \label{eq:traces}
\end{align}
with the cubic term $\varepsilon_{ij}M_{ij}$ of $\tr u^2$ being dropped as explained at the beginning of this subsection. Inserting \eqref{eq:traces} into
\eqref{eq:Fd2} and keeping terms quadratic and quartic in $\phi$, we get \eqref{eq_S_2d}
\begin{align}
  S_{d=2}=\chi \int \dd\tau\,\dd^2x\Big[
  &\tfrac{1 }{2}\big(\partial_\tau\partial_i\phi\big)^2
  +\tfrac{  c_T^2}{2}\big(\partial_i\partial_j\phi\big)^2
  +\tfrac{K}{2}\big(\partial_i\nabla^2\phi\big)^2+\lambda_1\,\big(\nabla^2\phi\big)^4 +\lambda_2\Big(\big(\partial_i\partial_j\phi\big)^2\Big)^2
  +\lambda_3\,\big(\nabla^2\phi\big)^2\big(\partial_i\partial_j\phi\big)^2
  \Big],
  \label{eq:Spsi}
\end{align}
where $(\partial_i\partial_j\phi)^2$ has been used in place of $(\nabla^2\phi)^2$
in the quadratic term, the two being equal under the integral, with the following identification of the 
couplings in the old basis,
\begin{align}
 \chi \lambda_1&=\tfrac18\sigma+\tfrac14w_3, \qquad \chi \lambda_2=\left(\tfrac{\chi c_L^2}{8}+\tfrac12\nu_2+w_3\right), \qquad
 \chi \lambda_3=-\tfrac14\sigma-\tfrac14\nu_2-w_3.
  \label{eq:matching}
\end{align}
The quartic couplings follow from the same expansion. The four invariants of
\eqref{eq:Fd2} that reach fourth order are $\lambda$, $\mu$, $\nu_2$ and $w_3$,
and they feed the three structures of \eqref{eq:Spsi}, so one combination is
redundant.
Further more,  by \eqref{eq:speeds}, $ c_T=0$  sets $\mu=-\sigma$ and we replace $\lambda$
in favour of $\chi c^2_L$ and $\mu$.
So at the floppy point there are exactly three independent marginal couplings. 

\section{Details of the RG in $d=4-\epsilon$}\label{app_RG_d4}

We study the IR fate of slow transverse phonons near $d=4$. We found that they were described by the Euclidean action \eqref{eq_S_4d}, which we copy here:
\begin{equation}\label{eq_S_4d_app}
\begin{split}
S = \int d\tau d^d x \, \frac12 & u_i^T \left(-\partial_\tau^2 - c_T^2 \nabla^2 + K \nabla^4\right)u_i^T + g \partial_i u^T_j \partial_i u^T_k \partial_j u^T_k\, .
\end{split}
\end{equation}
Importantly, the phonon propagator $G_{ij}(p) \equiv  \langle u^T_i u^T_j\rangle(p)$ includes a projector transverse to the momentum \eqref{eq_u_propagator}. We will study the one-loop renormalization of the couplings. By dimensional analysis, $K\sim \omega^2/q^4$, $u\sim (q^d/\omega)^{1/2}$ and $g\sim Kq/u\sim K^{5/4}  q^{(4-d)/2}$, so that $g$ is marginal at tree-level in $d=4$. At 1-loop, the running of couplings will take the form
\begin{equation}
\frac{d \log K}{d \log \Lambda}
	= \gamma_K \frac{g^2}{K^{5/2}} \, , \qquad
\frac{d \log c_\perp^2}{d \log \Lambda}
	= \gamma_c \frac{g^2}{K^{5/2}} \, , \qquad
\frac{d \log g}{d \log \Lambda}
	= \gamma_g \frac{g^2}{K^{5/2}} \, ,
\end{equation}
where $\gamma_K,\,\gamma_c,\,\gamma_g$ are dimensionless numbers that we will compute. As discussed in the main text, the kinetic term $(\partial_\tau u)^2$ remains unrenormalized at all orders in loops. The beta function for the dimensionless coupling $\hat g\equiv g/K^{5/4}$ is
\begin{equation}
\beta_{\hat g} 
	= \frac{d \hat g}{d \log \frac1\Lambda}
	\simeq \left(\frac54 \gamma_K - \gamma_g\right) \hat g^3\, ,
\end{equation}
and can thus feature a competition between two effects: renormalization of the propagator and of the vertex. We will find that the former wins, and $\beta_{\hat g} \geq 0$. It is thus possible to construct a perturbative fixed point in spatial dimention $d= 4-\epsilon$. The coupling $\hat g_c$ at the fixed point is
\begin{equation}
\beta_{\hat g_c}
	\simeq -\frac{\epsilon}{2} \hat g_c + \left(\frac54 \gamma_K - \gamma_g\right) \hat g_c^3 = 0 \qquad \Rightarrow \qquad
	\hat g_c^2 =  \frac{2\epsilon}{5\gamma_K - 4 \gamma_g}\, .
\end{equation}
At the fixed point, the dynamic critical exponent $z$ and correlation length exponent $\nu$ are given by
\begin{equation}
z \simeq 2 - \frac12 \gamma_K \hat g_c^2\, , \qquad
\nu \simeq \frac12 \left[ 1 +\frac12 (\gamma_K - \gamma_c) \hat g_c^2\right]\,,
\end{equation}
and the dimension of the relevant tuning operator is $\Delta_{\partial u^2} = d+z - \frac1\nu = d - \frac12 \gamma_K$. The rest of this appendix is devoted to the computation of $\gamma_K,\,\gamma_c,$ and $\gamma_g$.

\subsection{One-loop renormalization of the propagator}

The one-loop correction to the self-energy should include a leading UV-divergent contribution to $\delta c_\perp \sim \Lambda^2$, where $\Lambda$ is the UV cutoff on momentum $|\vb q|< \Lambda$, which we ignore because we are fine-tuning this term to zero. It should also include a subleading renormalization of the Lifshitz self-energy $\delta K \sim \log \Lambda$. Expanding out the path integral, the 1-loop correction to the Gaussian action is
\begin{equation}\label{eq_dSE}
-\delta S_E
	= \frac{g^2}{2} \int \partial_i u_j \partial_{i'} u_{j'} M^{iji'j'}\, , 
\end{equation}
(we omit the superscript for the transverse phonons $u^T$), with 
\begin{equation}\label{eq_M_2}
\begin{split}
M^{iji'j'}(p)
	= \langle (\partial_a u_b\partial_c u_d)(\partial_{a'} u_{b'}\partial_{c'} u_{d'})\rangle 
	&\left(\delta^{bd}\delta^{a}_{i}\delta^{c}_{j} + \delta^{bc}\delta^{a}_{i}\delta^{d}_{j} + \delta^{ac}\delta^{b}_{i}\delta^{d}_{j}\right)\\
	&\left(\delta^{b'd'}\delta^{a'}_{i'}\delta^{c'}_{j'} + \delta^{b'c'}\delta^{a'}_{i'}\delta^{d'}_{j'} + \delta^{a'c'}\delta^{b'}_{i'}\delta^{d'}_{j'}\right)\, , 
\end{split}
\end{equation}
where $(\cdot)$ denotes normal ordering. There are two possible Wick contractions:
\begin{equation}\label{eq_Wick}
\langle (\partial_a u_b\partial_c u_d)(\partial_{a'} u_{b'}\partial_{c'} u_{d'})\rangle 
	= \int_q q_a (p-q)_c q_{a'} (p-q)_{c'} G_{bb'}(q) G_{dd'}(p-q) + \left( ab \leftrightarrow cd\right),
\end{equation}
where $G_{ij} = \langle u_i u_j\rangle$ is the transverse phonon propagator \eqref{eq_u_propagator} (with $c_T^2 = 0$). 
We compute the $q_0$ integral by residue:
\begin{equation}\label{eq_residue}
\begin{split}
\int \frac{dq_0}{2\pi} \frac{1}{q_0^2 + \vb q^4}\frac{1}{(p_0-q_0)^2 + (\vb p - \vb q)^4}
	&= 
	\frac{1}{\vb q^2} \frac{1/2}{(p_0 - i \vb q^2)^2 + (\vb p - \vb q)^4} + 
	\frac{1}{(\vb p - \vb q)^2} \frac{1/2}{(p_0 + i \vb (\vb p - \vb q)^2)^2 + ( \vb q)^4} \\
	&\simeq \frac{1}{4\vb q^6} \left(1 + \frac{3 \vb p\cdot \vb q}{\vb q^2} + \frac{14 (\vb p\cdot \vb q)^2 - 3 \vb q^2 \vb p^2}{2 \vb q^4} + \cdots \right)
\end{split}
\end{equation}
where we set $K=1$ (it can be restored by dimensional analysis), and use the short-hand notation $\vb k^{2n} \equiv (\vb k \cdot \vb k)^n$. In the second line we expanded in $p$; there is no $\frac{p_0}{\vb q^2}$ term in the bracket, since the expression is even in $p_0$ (such a term would have been a total derivative later anyway).

The expression that must be integrated over spatial momenta $\int_{\vb q}\equiv \int \frac{d^d\vb q}{(2\pi)^d}$ is \eqref{eq_residue} multiplied by the index contraction
\begin{equation}
\begin{split}
\left[ q_a (p-q)_c q_{a'} (p-q)_{c'} P_{bb'}(q) P_{dd'}(p-q) + \left( ab \leftrightarrow cd\right)\right]
	&\left(\delta^{bd}\delta^{a}_{i}\delta^{c}_{j} + \delta^{bc}\delta^{a}_{i}\delta^{d}_{j} + \delta^{ac}\delta^{b}_{i}\delta^{d}_{j}\right)\\
	&\left(\delta^{b'd'}\delta^{a'}_{i'}\delta^{c'}_{j'} + \delta^{b'c'}\delta^{a'}_{i'}\delta^{d'}_{j'} + \delta^{a'c'}\delta^{b'}_{i'}\delta^{d'}_{j'}\right)\, .
\end{split}
\end{equation}
Dropping terms that involve longitudinal fields (e.g., terms with $\delta_{ij}$ or $p_i$) and using isotropy of the $\vb q$-integral, one finds
\begin{equation}\label{eq_a123}
M^{iji'j'}
	= \int_{\vb q}\delta^{ii'}\delta^{jj'} \frac{1}{\vb q^2} \left(a_1 + a_2 \frac{\vb p^2}{\vb q^2}\right) + a_3 \delta^{jj'} \frac{p_i p_{i'}}{\vb q^4}\, ,
\end{equation}
with coefficients $a_{1,2,3}$ given by
\begin{equation}
a_1 = -\frac{(d+4)(d-2)}{4d(d+2)} \to \frac16\, , \quad
a_2 = -\frac{(d+6)(d-2)(3d-2)}{8d(d+2)(d+4)} \to -\frac{25}{192}\, , \quad
a_3 = -\frac{(d-2)(2d+1)}{d(d+2)(d+4)}	\to -\frac{3}{32}\, .
\end{equation}
As a consistency check, notice that they are proportional to $d-2$: in $d=2$, there is no cubic interaction for purely transverse modes. The first $a_1$ is the power-law UV divergence absorbed by $c_\perp$; the other two will determine the renormalization of $K$.
Returning to \eqref{eq_dSE} and using $\int_{\vb q} \frac{1}{\vb q^4} = \frac{1}{8\pi^2}\log \Lambda$ in $d=4$, we find
\begin{equation}
\delta S_E = - \frac{g^2}{2} \int (\nabla^2 u)^2 (a_2 + a_3) \frac{1}{8 \pi^2} \log \Lambda
\end{equation}
so that $\delta K = \frac{g^2}{8\pi^2} (-a_2-a_3)$ or 
\begin{equation}
\gamma_K = - \frac{a_2 + a_3}{8\pi^2} 
	= \frac{43}{192} \frac{1}{8\pi^2}\, .
\end{equation}
%

\subsection{Correlation length critical exponent}

The correlation length critical exponent $\nu$, related to the renormalization of the tuning parameter $c_\perp^2$, can be obtained from the calculations above with a small modification: one insertion of the relevant operator in the loops, as shows below.

\centerline{  
\begin{tikzpicture}[scale=1, line width=0.8pt]
    \coordinate (v1) at (-0.6, 0);
    \coordinate (v2) at (0.6, 0);

    \draw (v1) -- (-1.2, 0);
    \draw (v2) -- (1.2, 0);

    \draw (v1) to[out=60, in=120] (v2);
    \draw (v1) to[out=-60, in=-120] (v2);

    \fill (0,0.3) circle (2.5pt);
\end{tikzpicture}
}

\noindent
Eq.~\eqref{eq_Wick} still holds, but we now use the phonon dispersion $E(q)^2 = c_T^2 q^2 + K q^4$, and keep only the term linear in $c_T^2$, so that the $q_0$ integral becomes
\begin{equation}
\int \frac{dq_0}{2\pi} \frac{1}{(q_0^2 + E(\vb q)^2)^2} \supset -2c_T^2 \int \frac{dq_0}{2\pi} \frac{\vb q^2}{(q_0^2 + K \vb q^4)^3}
	= -\frac{3c_T^2}{8K^{5/2}} \frac{1}{\vb q^8}\, , 
\end{equation}
where in the second step we only kept the linear term in $c_T^2$. Note that we can set the external momentum $p\to 0$, since the leading UV divergence is now logarithmic. This now leads to an additional $a_1$ term in \eqref{eq_a123}:
\begin{equation}
M^{iji'j'}|_{c_T^2}
	= -\frac{3c_T^2}{2K^{5/2}}  \delta^{ii'}\delta^{jj'} \int_{\vb q} \frac{1}{\vb q^4} a_1
	= -\frac{c_T^2}{4K^{5/2}}  \delta^{ii'}\delta^{jj'} \frac{\log \Lambda}{8\pi^2}\, .
\end{equation}
This generates a correction to the parameter $\delta c_T^2 = c_T^2\hat g^2 \frac14 \frac{1}{8\pi^2} \log \Lambda$, or
\begin{equation}
\gamma_c = \frac{1}{4} \frac{1}{8\pi^2}\, .
\end{equation}
%

\subsection{One-loop renormalization of the vertex}
A vertex correction also gets generated at one-loop:
\begin{equation}
\delta S_E
	= \frac{g^3}{3!} \int \partial_i u_j\partial_{i'} u_{j'}\partial_{i''} u_{j''} M_{iji'j'i''j''}
\end{equation}
with
\begin{equation}
M_{iji'j'i''j''}
	= \langle (\partial_a u_b\partial_c u_d)(\partial_{a'} u_{b'}\partial_{c'} u_{d'})(\partial_{a''} u_{b''}\partial_{c''} u_{d''})\rangle 
	\left(\delta^{bd}\delta^{a}_{i}\delta^{c}_{j} + \delta^{bc}\delta^{a}_{i}\delta^{d}_{j} + \delta^{ac}\delta^{b}_{i}\delta^{d}_{j}\right)()'()''\, , 
\end{equation}
where $()'$ and $()''$ denote the same Kronecker deltas with primed indices, similar to \eqref{eq_M_2}. There are eight possible Wick contractions:
\begin{equation}
\begin{split}
&\langle (\partial_a u_b\partial_c u_d)(\partial_{a'} u_{b'}\partial_{c'} u_{d'})(\partial_{a''} u_{b''}\partial_{c''} u_{d''})\rangle \\
&= \left(2\int_q q_a q_{a'}(p_1 + q)_{c'}(p_1 + q)_{c''}(p_2-q)_c(p_2-q)_{a''}G_{bb'}(q) G_{d'd''}(p_1 + q)G_{b''d}(p_2 - q)\right.\\
&\left.\qquad\quad + (a'b' \leftrightarrow c'd')\right) + (a''b'' \leftrightarrow c''d'') \, .
\end{split}
\end{equation}
%
In this calculation, we are only interested in the leading UV divergence, which is logarithmic. We can therefore set the external momenta $p_1,p_2\to 0$. The integral over frequency can be obtained by residue
\begin{equation}\label{eq_residue_3}
\int \frac{dq_0}{2\pi}
	\left(\frac{1}{q_0^2 + \vb q^4}\right)^3 = \frac{3}{16} \frac{1}{\vb q^{10}}\, .
\end{equation}
The expression that must be integrated over $\int_{\vb q}$ is thus \eqref{eq_residue_3} times
\begin{equation}
\begin{split}
&2\Bigl[ 
q_a q_{a'} q_{a''}q_c q_{c'} q_{c''}P_{bb'}(q) P_{d'd''}(q)P_{b''d}(q)\\
&+ (a'b' \leftrightarrow c'd') + (a''b'' \leftrightarrow c''d'')\Bigr]\left(\delta^{bd}\delta^{a}_{i}\delta^{c}_{j} + \delta^{bc}\delta^{a}_{i}\delta^{d}_{j} + \delta^{ac}\delta^{b}_{i}\delta^{d}_{j}\right)()'()''\, .
\end{split}
\end{equation}
One finds:
\begin{equation}
\delta S_E 
	= \frac{g^3}{3!} \int \partial_i u_j \partial_i u_k \partial_j u_k \frac{b_1 }{8\pi^2} \log \Lambda\, , \qquad \hbox{with} \quad 
	b_1 = 9 \frac{d^3 + 3d^2 - 8 d - 64}{8d(d+2)(d+4)} \to \frac{3}{32} \, .
\end{equation}
So the radiatively generated coupling is  $\delta g = \frac{g^2}{3!} \frac{3}{32} \frac{\log \Lambda}{8\pi^2} = \frac{g^2}{64} \frac{\log \Lambda}{8\pi^2}$, or
\begin{equation}
\gamma_g = \frac{1}{64} \frac{1}{8\pi^2}.
\end{equation}
%

\section{Observables at the crystal Lifshitz QCP}\label{app_obs}

The arguments in the main text establish the existence of a strongly coupled Lifshitz QCP describing ``floppy'' quantum solids, with vanishing transverse phonon velocities. In this appendix, we cover a number of experimentally accessible observables that probe this crystal Lifshitz universality class.

\subsection{Momentum response}
The phonon propagators at the fixed point are
\begin{equation}\label{eq_uu}
\langle u_i u_j\rangle(\omega,q)
	\simeq \frac{1}{\chi} \left[\frac{q_iq_j}{q^2} \frac{1}{c_L^2 q^2 - \omega^2} + \left(\delta_{ij} - \frac{q_iq_j}{q^2}\right) \frac1{\omega^2} F(\omega/q^z) \right]\, ,
\end{equation}
where $F$ is a universal scaling function of the interacting Lifshitz QCP. The fact that the transverse sector scales as $1/\omega^2 $ or $1/q^{2z}$, and does not involve an anomalous dimension (other than $z$) follows from the nonrenormalization of the kinetic term $\chi \dot u^2$ (or momentum susceptibility) to all orders in perturbation theory. The universal scaling function can in principle be computed within our $\epsilon$-expansion. Some aspects of it can be determined non-perturbatively: $F(w\to \infty)=-1$ follows from the nonrenormalization of $\chi$, and $F(w) \sim w^2$ as $w^2\to 0$ for the correlator to be finite at zero frequency. At the free (Gaussian) Lifshitz fixed point ($z=2$), it is given by $F(w) = \frac{w^2}{1-w^2}$.

This leads to a similar structure for the two-point function of the momentum density operator $T_{0i} = \chi \dot u_i + \cdots$. To leading order at small $\omega,q$:
\begin{equation}\label{eq_TT}
\langle T_{0i} T_{0j}\rangle(\omega,q)
	\simeq \chi \left[\frac{q_iq_j}{q^2} \frac{c_L^2 q^2}{c_L^2 q^2 - \omega^2} + \left(\delta_{ij} - \frac{q_iq_j}{q^2}\right) \tilde F(\omega/q^z) \right]\, ,
\end{equation}
which can be obtained by \eqref{eq_uu} by multiplying with $\omega^2$ and adding an analytic (in fact, constant) contact term to guarantee that the static limit gives the static momentum susceptibility $\lim_{k\to 0}\lim_{\omega\to 0} \langle T_{0i} T_{0j}\rangle(\omega,q) = \chi \delta_{ij}$. Thus, the universal scaling function appearing here is related to the previous one as $\tilde F(w) = F(w) + 1$.

\subsection{Density response}

Consider the density operator $\rho(t,\vb x) = \sum_n \delta^d(\vb x - \vb R_n(t))$, where $\vb R_n(t) = \bar {\vb R}_n + \vb u_n(t)$ is the location of atom $n$ (Lagrangian description), and $\bar {\vb R}_n$ denotes the reference lattice. In the continuum limit, we define the map $\vb R(t,\vb x)$ such that $\vb R(t,\bar{\vb R}_n) = \vb R_n(t)$. It is useful to also define the inverse map $\vb X = \vb R^{-1}$ (Eulerian specification), such that $\vb X (\vb R(\vb x)) = \vb x$ (dropping the time argument $t$ for readability). With these definitions, one can obtain an expression for the density operator in the EFT: 
\begin{equation}
\rho(t,\vb x)
	= \sum_{n} \delta^d(\vb x - \vb R(t,\bar{\vb R}_n)))
	= \sum_{n} \left|\det \frac{\partial \vb X(t,\vb x)}{\partial \vb x}\right| \delta^d(\vb X(\vb x) - \bar{\vb R}_n)
	= \rho_0 \left|\det \frac{\partial \vb X(t,\vb x)}{\partial \vb x}\right| \sum_{\vb G} e^{i\vb G \cdot \vb X(t,\vb x)} \, .
\end{equation}
In the last step, we used the Poisson resummation formula to convert the sum over the lattice into a sum over reciprocal lattice vectors $\vb G$ (satisfying $\vb G \cdot \bar{\vb R}_i\in 2\pi \mathbb Z$), and $\rho_0$ is the density of the lattice. This expression can also be written as
\begin{equation}
\begin{split}
\rho(t,\vb x)
	&= \rho_0 \frac1{d!}\epsilon^{i_1 \cdots i_d}\epsilon_{j_1\cdots j_d} \partial_{i_1} X^{j^1}\cdots \partial_{i_d} X^{j^d}\sum_{\vb G} e^{i\vb G \cdot \vb X(t,\vb x)}\, .
\end{split}
\end{equation}
At wavelengths much larger than lattice constants, only the $\vb G = 0$ term in the sum contributes. Expanding $\vb X(t,\vb x) = \vb x - \vb u(t,\vb X(t,\vb x))$ to linearized order, one finds that the density only involves the longitudinal phonon $\rho\propto \nabla \cdot u$. Thus, at small wavevectors, density response only receives subleading contributions from the transverse sector. This will involve any scalar operator of the transverse sector; the scalar with the lowest scaling dimension is the composite operator $(\nabla u_T)^2$, which determines the exponent $\nu$.

Since we are precisely interested in the transverse piece, we turn to density response near a reciprocal lattice vector: $\vb q = \vb G + \vb k$, $k\ll G$, which we will find is dominated by the strongly coupled transverse sector. To leading order, the density is now given by a vertex operator $\rho(t,\vb x) \simeq \rho_0 e^{i\vb G\cdot \vb X(t,\vb x)}$. In general, the dimension of this operator is independent from that of the phonon field $u$. However, at weak coupling near the upper critical dimension $\epsilon = 4- d \ll 1$, we can expand in fluctuations $u$ and find that the two are related. First, at the Gaussian Lifshitz fixed point one has, for density response near a wavevector $\vb G$:
\begin{equation}
\langle \rho(t,\vb x) \rho(0,0)\rangle
	= \rho_0^2 \langle e^{i \vb G \cdot \vb X(t,\vb x)}e^{-i \vb G \cdot \vb X(0,0)}\rangle
	\simeq \rho_0^2 e^{i \vb G \cdot \vb x} e^{-\frac12 \langle (\vb G\cdot( \vb u(t,\vb x) - \vb u(0,0) )^2  \rangle}
	= \rho_0^2 e^{i\vb G\cdot \vb x} e^{- G^2 \langle u^2\rangle} 
	e^{G^i G^j\langle u_i(t,\vb x) u_j\rangle}\, .
\end{equation}
where in the second step we expanded $\vb X(t,\vb x) = \vb x - \vb u(t,\vb X(t,\vb x))$ in fluctuations $u$.
The spacetime independent factor $e^{- G^2 \langle u^2\rangle}$ that arose from normal ordering the vertex operators is referred to as the Debye-Waller factor in the context of elasticity. Crucially, it is only IR finite ($\int_0^\Lambda \frac{d\omega d^dq}{\omega^2 + q^4}<\infty$) in spatial dimension $d>2$. In $d=2$, quantum critical solids feature an interesting spectrum of vertex operators with dimension $\Delta = \frac{Q^2}{8\pi \chi \sqrt{K}}$, with the dimensionless parameter $Q^2 / (\chi \sqrt{K})$ playing a similar role to the Luttinger parameter (or radius) of the $d=1$ compact scalar. A similar phenomenon occurs in the $d=2$ scalar Lifshitz QCP \cite{Vishwanath:2003yjl}. Assuming in what follows that $d>2$, 
\begin{equation}
\langle \rho(\omega, \vb G+\vb k) \rho(0,0)\rangle
	\simeq \rho_0^2 e^{-G^2 \langle u^2\rangle}
	\left((2\pi)^{d+1} \delta(\omega) \delta^d(\vb k) + G^i G^j \langle u_i u_j\rangle(\omega,\vb k)\right)\, .
\end{equation}
The first term is the Bragg peak of the equal time structure factor, whereas the second term is the contribution from the fluctuating phonons, excited at low momentum $k\ll G$. Using \eqref{eq_uu}, we find that the low frequency $0<\omega \lesssim k^z$ behavior is dominated by the critical transverse phonons:
\begin{equation}
S(\omega,\vb G + \vb k)
	\simeq \rho_0^2 e^{-G^2 \langle u^2\rangle} G^2(1-\cos^2 \phi) \frac1{\omega^2} F(\omega/q^z)\, , 
\end{equation}
where $\cos \phi = \vb G \cdot \vb k/|Gk|$, and we only kept the transverse contribution which dominates at low frequencies $\omega\lesssim k^z$, for generic angles $\phi$. Note that while the nonuniversal prefactor depends on the reciprocal lattice vector $\vb G$, it is independent of $\omega$ or $q$.

We close our discussion of the charge density operator with a more formal comment. It is somewhat surprising that the elasticity theory has a conserved (number) density; after all, this EFT is associated with the spontaneous breaking of spatial translation symmetry, and could apply to systems with or without a microscopic $U(1)$ symmetry. In fact, the $U(1)$ symmetry of the elasticity EFT should be viewed as an emergent (or accidental) symmetry, carried by the current
\begin{equation}
j^\mu = \rho_0 \frac1{d!}\epsilon^{\mu\mu_1\cdots \mu_d} \epsilon_{i_1 \cdots i_d} \partial_{\mu_1} X^{i_1} \cdots \partial_{\mu_d} X^{i_d} \sum_{\vb G} e^{i\vb G \cdot \vb X(t,\vb x)}\, .
\end{equation}
The low-momentum limit of this expression (where the sum $\sum_{\vb G}$ can be dropped) has been appreciated for some time \cite{souriau1959matiere,Son:2005ak}; however, we emphasize that correlators of this operator near Bragg peaks $\vb q \simeq \vb Q$ are also captured by the low energy EFT. Now, when the underlying system has a {\em microscopic} $U(1)$ symmetry, the two symmetries can be matched through an argument similar to the one that identifies the skyrmion number and baryon number in the chiral Lagrangian describing pions \cite{Witten:1983tw}: in the present case, this can be done by coupling the $U(1)$ current to a background field $A_\mu$, and comparing the Lorentz force $\partial_\mu T^{\mu i} = F^{i\nu} j_\nu$ between microscopics and EFT.

\subsection{Nonlinear elasticity and thermal dynamics}

The shear modulus vanishes at the crystal Lifshitz fixed point:
\begin{equation}
\mu \equiv \lim_{\omega\to 0}\lim_{k\to 0} \langle T_{xy} T_{xy}\rangle(\omega,k) = 0\,.
\end{equation}
This leads to a nonlinear response to shear strain. The correlator \label{eq_TT} implies that quantum fluctuations of momentum density scale as $\delta T_{0i} \sim (\omega q^d)^{1/2}$, so that stress tensor fluctuations scale as $\delta T_{ij} \sim \frac{\omega}{q}(\omega q^d)^{1/2}$. The response to a change in the background metric $\delta S  = \int dt d^dx \, \delta g_{ij} T_{ij}$ thus takes the form (using $\delta S \sim 1 \Rightarrow \delta g \sim \frac{q}{\omega}(\omega q^d)^{1/2}$ )
\begin{equation}
\langle \delta T_{xy}\rangle
	\sim (\delta g_{xy})^{\delta}\, , \qquad
	\delta  = \frac{d+3z - 2}{d + 2 - z}\, .
\end{equation}

In the context of elasticity, it is more common to study the {\em strain} $\epsilon = \langle u_{xy}\rangle \simeq \langle \partial_x u_y\rangle$ in response to an external stress $\sigma$, which then involves the inverse power $1/\delta$:
\begin{equation}\label{eq_delta}
\epsilon(\sigma,c_T^2) \sim \sigma^{1/\delta}  {\hat f_\epsilon}(c_T^{\nu(d+3z-2)}/ \sigma)\, .
\end{equation}
We recover linear response and Hooke's law for regular phonons ($z=1$), but find nonlinear response ($\delta>1$) at the Lifshitz QCPs ($z>1$). At the upper critical dimensions for quartic couplings ($d=2$) and cubic couplings ($d=4$), where $z=2$, the exponent reduces to its mean field values $\delta = 3$ and $\delta =2$ respectively. In the $\epsilon = 4-d$ expansion, it is given by $\delta = 2- \frac{3}{203}\epsilon$. 

Linear response is recovered if one tunes away from criticality, $c_T^2 \neq 0$. This is captured by the universal scaling function $F_\epsilon$ above, which describes the crossover between nonlinear response at criticality ($F_{\epsilon}(0) = 1 \Rightarrow \epsilon\sim \sigma^{1/\delta}$), and linear response slightly away from criticality, albeit with a singular dependence on the tuning parameter: $\epsilon \sim \sigma \times (c_T^2)^{-2(z-1)}$ (which implies the limiting behavior $\lim_{s\to \infty} F(s) \propto 1/s^{1-\frac1\delta}$).

The scaling argument above implies that the shear modulus scales as $\mu \sim q^{2(z-1)}$ and thus indeed vanishes as $q\to 0$ at the Lifshitz fixed point $z>1$. A nonzero but singular shear modulus is obtained either at finite temperature $T\neq 0$, or detuning away from the QCP $c_T^2 \neq 0$:
\begin{equation}\label{eq_mu_scale}
\mu \sim T^{2(z-1)/z}\, , \qquad
\mu \sim (c_T^2)^{2\nu(z-1)}\, .
\end{equation}
Equations \eqref{eq_delta} and \eqref{eq_mu_scale} can be summarized with the following scaling form for the free energy density or pressure
\begin{equation}
P(T,c_T^2,\sigma)
	\sim T^{\frac{d}{z} + 1} \mathcal F \left(\frac{c_T^2}{T^{1/\nu z}}, \frac{\sigma^2}{T^{(d-2+3z)/z}}\right)\, .
\end{equation}
Taking one derivative with respect to $\sigma$ and setting $T$ to zero produces \eqref{eq_delta}, while taking a second derivative and setting instead $\sigma=0$ gives \eqref{eq_mu_scale}. This expression also encodes the anomalous specific heat of floppy solids in the quantum critical fan:
\begin{equation}
C_V \sim T\frac{d^2 P}{dT^2}
	\sim T^{d/z}\, .
\end{equation}

Finally, we comment briefly on dynamics at finite temperature, in the quantum critical fan above the Lifshitz QCP. At high frequencies $\omega \gg T$, the system has not thermalized and correlation functions are still those of the Lifshitz QCP shown above. In the opposite limit $\omega\ll T$, dynamics is described by the dissipative hydrodynamics of solids \cite{ChaikinLubensky}, albeit with unusual temperature dependence of transport parameters reflecting the underlying QCP. For example, the dynamical shear modulus $\mu(\omega)$ or shear viscosity $\eta(\omega)$ have the scaling behavior
\begin{equation}
\eta(\omega,T)
	 = \frac{\mu(\omega,T)}{-i\omega}
	 \sim \frac{\omega^{2(z-1)/z}}{-i\omega} \hat f_\mu(\omega/T)\, , 
\end{equation}
with $\hat f_\eta$ a universal scaling function of the crystal Lifshitz QCP with the limiting behavior $\hat f_\mu(\infty) = 1$, and $\hat f_\mu (x\to 0) \propto 1/x^{2(z-1)/z}$. In particular, at $T=0$ the dynamic shear modulus scales as $\mu(\omega,T=0)\sim \omega^{2(z-1)/z}$.


\section{Details of the RG in $d=2$}\label{app_RG_d2}

We now examine the infrared behavior of the floppy theory near two spatial dimensions. As shown in Sec.~\ref{app_EFT_IC}, in exactly two dimensions the transverse displacement can be expressed as $u_i^T=\epsilon_{ij}\partial_j\phi$. In this preliminary study, we restrict the interaction sector to transverse fluctuations by setting $u_i^L=0$, leaving a systematic treatment of longitudinal–transverse mixing to future work. The purely transverse cubic interaction then vanishes identically in $d=2$, and the leading interactions within this truncation are quartic.
In the conventions of Sec.~\eqref{app_EFT}, the Euclidean action is
\begin{align}
S_E^{d=2}=\int \mathrm d\tau\,\mathrm d^2x\,
\Bigg[&\frac12\bigl(\partial_\tau\partial_i\phi\bigr)^2
+\frac{c_T^2}{2}\bigl(\partial_i\partial_j\phi\bigr)^2
+\frac K2\bigl(\partial_i\nabla^2\phi\bigr)^2
\nonumber\\
&+\lambda_1\bigl(\nabla^2\phi\bigr)^4
+\lambda_2\left(\bigl(\partial_i\partial_j\phi\bigr)^2\right)^2
+\lambda_3\bigl(\nabla^2\phi\bigr)^2
              \bigl(\partial_i\partial_j\phi\bigr)^2\Bigg].
\label{eq:d2-action}
\end{align}
The couplings are related to those of Sec.~\eqref{app_EFT} by
\begin{align}
\lambda_1&=\frac18\sigma+\frac14w_3,
&
\lambda_2&=\frac{c_L^2}{8}+\frac12\nu_2+w_3,
&
\lambda_3&=-\frac14\sigma-\frac14\nu_2-w_3.
\label{eq:d2-coupling-map-sigma}
\end{align}
As before, the redundancy among the quartic structures may be used to remove
\(\sigma\).  We choose the same representative as in Sec.~\eqref{app_EFT},
\begin{align}
\lambda_1&\simeq\frac14w_3,
&
\lambda_2&\simeq\frac{c_L^2}{8}+\frac12\nu_2+w_3,
&
\lambda_3&\simeq-\frac14\nu_2-w_3.
\label{eq:d2-coupling-map}
\end{align}
At the floppy point, \(c_T^2=0\), the scalar propagator is,
\begin{align}
s_p^\phi\equiv\langle\phi(p)\phi(q)\rangle
=\frac{\delta^{(3)}(p+q)}{\bigl(\omega_p^2+K\vb p^{\,4}\bigr)\vb p^{\,2}}.
\label{eq:d2-propagator}
\end{align}

With the \(z=2\) scaling of the Gaussian theory, the quartic couplings are
marginal in \(d=2\).  The logarithmic renormalization of these couplings is
therefore determined by two insertions of the quartic interaction.  At the
same order a one-vertex tadpole can shift the relevant quadratic coupling by
a power divergence, but it does not produce the logarithmic correction to
\(K\) considered here.  We consequently keep \(K\) fixed throughout this
calculation.

\subsection{One-loop renormalization of the quartic vertices}

We evaluate the correction directly at the level of the Euclidean action.
Writing the Hessian as a slowly varying background plus a fluctuation, we
retain from each interaction the terms containing two background fields and
two fluctuating fields.  We denote the background Hessian by
\(H_{ij}=\partial_i\partial_j\bar\phi\), while, as in the calculation below,
the Fourier modes \(\phi_q\) denote the fluctuating field.  We also write
\(\int_{p,q}\) for the integrations over the two frequency-momenta.  The
three interaction terms then give
\begin{align}
\lambda_1\int_x\bigl(\nabla^2\phi\bigr)^4
&\simeq
6\lambda_1\int_{p,q}
\vb q^{\,2}(\vb p+\vb q)^{2}
\bigl(\operatorname{tr}H\bigr)^2(p)\,
\phi_q\phi_{-p-q},
\nonumber\\
\lambda_2\int_x
\left(\bigl(\partial_i\partial_j\phi\bigr)^2\right)^2
&\simeq
\lambda_2\int_{p,q}
\Bigl[
2\operatorname{tr}H^2(p)
 \bigl(\vb q\mathbin{\cdot}(\vb p+\vb q)\bigr)^2+4H_{ij}H_{kl}(p)q_iq_j(p+q)_k(p+q)_l
\Bigr]\phi_q\phi_{-p-q},
\nonumber\\
\lambda_3\int_x
\bigl(\nabla^2\phi\bigr)^2
\bigl(\partial_i\partial_j\phi\bigr)^2
&\simeq
\lambda_3\int_{p,q}
\Bigl[
\bigl(\operatorname{tr}H\bigr)^2(p)
 \bigl(\vb q\mathbin{\cdot}(\vb p+\vb q)\bigr)^2+4\operatorname{tr}H\,H_{ij}(p)\vb q^{\,2}
 (p+q)_i(p+q)_j
+\operatorname{tr}H^2(p)\vb q^{\,2}(\vb p+\vb q)^2
\Bigr]\phi_q\phi_{-p-q}.
\label{eq:d2-background-expansion}
\end{align}
The second cumulant therefore gives,
\begin{align}
\delta S_E
=-\frac12\int_{p,q,p',q'}
M(p,q)M(p',q')
\left\langle
\bigl(\phi_q\phi_{-p-q}\bigr)
\bigl(\phi_{q'}\phi_{-p'-q'}\bigr)
\right\rangle .
\label{eq:d2-cumulant}
\end{align}
Here \(M(p,q)\) is the coefficient of
\(\phi_q\phi_{-p-q}\) in Eq.~\eqref{eq:d2-background-expansion}, namely
\begin{align}
M(p,q)={}&
6\lambda_1\vb q^{\,2}(\vb p+\vb q)^2
 \bigl(\operatorname{tr}H\bigr)^2(p)
\nonumber\\
&+\lambda_2\Bigl[
2\operatorname{tr}H^2(p)
 \bigl(\vb q\mathbin{\cdot}(\vb p+\vb q)\bigr)^2
+4H_{ij}H_{kl}(p)q_iq_j(p+q)_k(p+q)_l
\Bigr]
\nonumber\\
&+\lambda_3\Bigl[
\bigl(\operatorname{tr}H\bigr)^2(p)
 \bigl(\vb q\mathbin{\cdot}(\vb p+\vb q)\bigr)^2
+4\operatorname{tr}H\,H_{ij}(p)\vb q^{\,2}
 (p+q)_i(p+q)_j
\nonumber\\[-0.2em]
&\hspace{2.55cm}
+\operatorname{tr}H^2(p)\vb q^{\,2}(\vb p+\vb q)^2
\Bigr].
\label{eq:d2-M-definition}
\end{align}
There are two connected Wick contractions.  In the first, the momentum delta
functions set \(p'=-p\) and \(q'=-q\); in the second they set \(p'=-p\) and
\(q'=p+q\).  Stripping the overall momentum-conserving delta function and
denoting the propagator without its delta function by \(S_q\), the result is
\begin{align}
\delta S_E
=-\frac12\int_{p,q}
\Bigl[
M(p,q)M(-p,-q)+M(p,q)M(-p,p+q)
\Bigr]S_qS_{-p-q}.
\label{eq:d2-two-contractions}
\end{align}

Eq.~\eqref{eq:d2-two-contractions} is the full momentum-dependent
expression.  To extract the renormalization of the operators already present
in the original action, we perform a local derivative expansion about $p=0$, with the higher powers of external momenta generating higher-derivative operators,
such as \((\partial_\mu H_{ij})^2H_{kl}H_{kl}\). The frequency integral required for the logarithmically divergent local term
is,
\begin{align}
\int\frac{\mathrm dq_0}{2\pi}\,S_qS_{-p-q}
\simeq\frac{1}{4K^{3/2}\vb q^{\,10}}+\cdots ,
\label{eq:d2-frequency-integral}
\end{align}
where $\cdots$ denote terms which are $O(\vb p^2)$. Carrying out the remaining tensor contractions and the isotropic angular
integral then gives
\begin{align}
\delta S_E=-\int_x\frac{\log b}{64\pi}
\Bigg\{&
\Bigl[
288\lambda_1^2-13\lambda_2^2-8\lambda_2\lambda_3
+56\lambda_3^2+48\lambda_1(\lambda_2+6\lambda_3)
\Bigr]\bigl(\operatorname{tr}H\bigr)^4
\nonumber\\
&+4\Bigl[
13\lambda_2^2+54\lambda_2\lambda_3+20\lambda_3^2
+24\lambda_1(3\lambda_2+\lambda_3)
\Bigr]
\bigl(\operatorname{tr}H\bigr)^2\operatorname{tr}H^2
\nonumber\\
&+4\Bigl[
19\lambda_2^2+12\lambda_2\lambda_3+2\lambda_3^2
\Bigr]\bigl(\operatorname{tr}H^2\bigr)^2
\Bigg\}.
\label{eq:d2-loop-divergence}
\end{align}
In order to get this result we have used Eq.~\eqref{eq:CH} since such higher traces get generated at one loop. As in Sec.~\eqref{app_RG_d4}, we have displayed \(K\) in the frequency integral and set
\(K=1\) in the final logarithmic coefficients.  Its dependence can be
restored by dimensional analysis. Following the convention used in this work, the cutoff is lowered toward the
infrared and the momentum integrals are performed by integrating out a momentum shell between $\Lambda$ and $\Lambda/b$. Using the map in Eq.~\eqref{eq:d2-coupling-map}, the
one-loop contributions at \(d=2\) are,
\begin{align}
-\frac{\delta c_L^2}{\delta\log[b]}
\equiv\beta_{c_L^2}
&=\frac{4c_L^4+4c_L^2\nu_2+\nu_2^2}{16\pi},
\label{eq:d2-beta-CL}
\\
-\frac{\delta\nu_2}{\delta\log[b]}
\equiv\beta_{\nu_2}
&=\frac{
32c_L^2w_3+23c_L^2\nu_2+16w_3\nu_2+16\nu_2^2
}{64\pi},
\label{eq:d2-beta-nu}
\\
-\frac{\delta w_3}{\delta\log[b]}
\equiv\beta_{w_3}
&=\frac{
-13c_L^4-48c_L^2w_3+576w_3^2-88c_L^2\nu_2
+576w_3\nu_2+80\nu_2^2
}{1024\pi}.
\label{eq:d2-beta-w}
\end{align}
Eqs.~\eqref{eq:d2-beta-CL}-\eqref{eq:d2-beta-w} are the logarithmic
loop contributions in exactly two dimensions. 

\subsection{Mean-field stability}

\begin{figure}
\centerline{
\subfloat[]{\label{fig:stabilityH2}
\includegraphics[width=0.45\linewidth, angle=0]{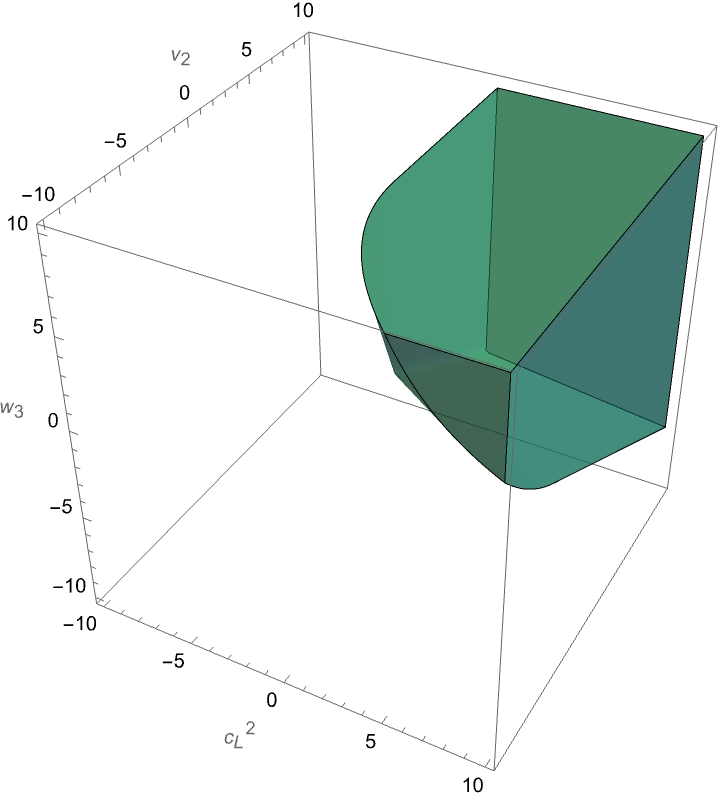}}
\subfloat[]{\label{fig:RGflows2D2}
\includegraphics[width=0.45\linewidth, angle=0]{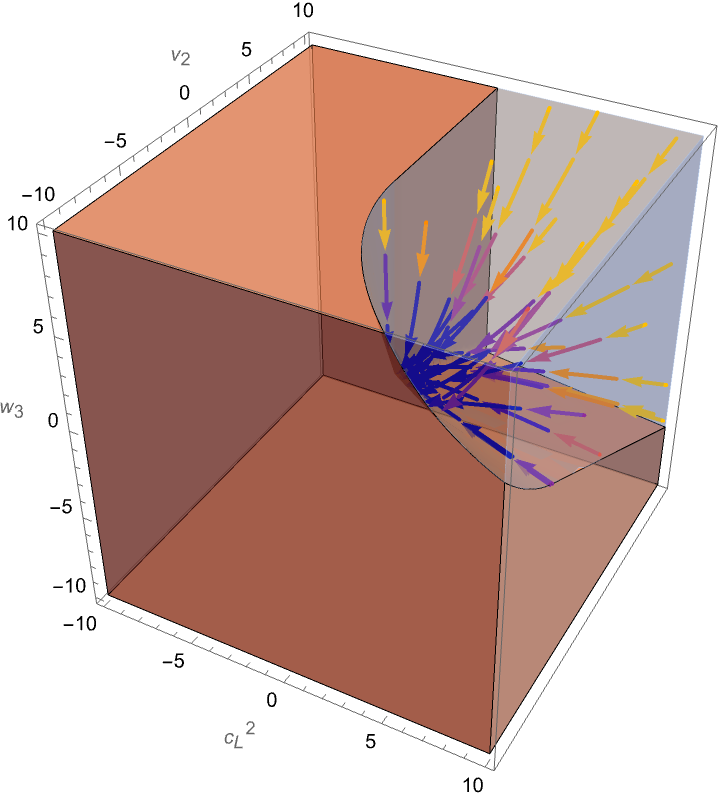}}
}\caption{(a) Stability region of the couplings $c_L^2, \nu_2$ and $w_3$ due to \eqref{eq:d2-endpoint-conditions} and \eqref{eq:d2-interior-condition}. (b) RG flows of the couplings  $c_L^2, \nu_2$ and $w_3$ along with the region excluded by stability conditions \eqref{eq:d2-endpoint-conditions} and \eqref{eq:d2-interior-condition}.}
\end{figure}

We finally determine the region in coupling space for which the quartic
potential is positive.  In terms of the two invariants of the real symmetric
\(2\times2\) Hessian,
\(T=\operatorname{tr}H\) and \(R=\operatorname{tr}(H^2)\), the potential is
\begin{align}
V(H)=\lambda_1T^4+\lambda_2R^2+\lambda_3T^2R.
\label{eq:d2-potential}
\end{align}
Let \(z_1,z_2\) be the eigenvalues of \(H\).  Using
Eq.~\eqref{eq:d2-coupling-map}, positivity is equivalent to
\begin{align}
c_L^2+8w_3(t-1)^2-4\nu_2(t-1)&\geq0,
\qquad 0\leq t\leq1,\qquad 
t=\frac{(\operatorname{tr}H)^2}{2\operatorname{tr}(H^2)}
=\frac{(z_1+z_2)^2}{2(z_1^2+z_2^2)}.
\label{eq:d2-stability-polynomial}
\end{align}
The endpoint conditions are
\begin{align}
c_L^2+8w_3+4\nu_2\geq0,
\qquad c_L^2\geq0.
\label{eq:d2-endpoint-conditions}
\end{align}
For \(w_3\leq0\), the polynomial is concave, so its minimum on the closed
interval occurs at an endpoint and these two conditions are sufficient.  For
\(w_3>0\), the stationary point lies at
\(t_\star=(4w_3+\nu_2)/(4w_3)\).  When \(0<t_\star<1\), one must impose the
additional condition
\begin{align}
c_L^2-\frac{\nu_2^2}{2w_3}\geq0,
\qquad
w_3>0,
\qquad
0\leq\frac{4w_3+\nu_2}{4w_3}\leq1.
\label{eq:d2-interior-condition}
\end{align}
This last inequality follows from the minimum of the parabola, not its
maximum. We plot the constraint region in Fig.~\ref{fig:stabilityH2} and the RG flows dictated by \eqref{eq:d2-beta-nu} in Fig.~\ref{fig:RGflows2D2}, where we also indicate the region forbidden by the stability analysis.

\end{document}